\documentclass[aps,preprint]{revtex4-1}
\usepackage{graphicx}
\usepackage{amsmath,amssymb,amsfonts}

\newcommand{\vR}{\mathbf{R}}
\newcommand{\vvr}{\mathbf{r}}

\begin{document}

\title{Effective model of protein--mediated interactions in chromatin}

\author{Francesco Borando}
\affiliation{Department of Physics, Universit\`a degli Studi di Milano, via Celoria 16, 20133 Milano, Italy}
\author{Guido Tiana}
\affiliation{Department of Physics and Center for Complexity and Biosystems, Universit\`a degli Studi di Milano and INFN, via Celoria 16, 20133 Milano, Italy}

\email{guido.tiana@unimi.it}
\date{\today}

\begin{abstract}
Protein--mediated interactions are ubiquitous in the cellular environment, and particularly in the nucleus, where they are responsible for the structuring of chromatin. We show through molecular--dynamics simulations of a polymer surrounded by binders that the strength of the binder--polymer interaction separates an equilibrium from a non--equilibrium regime. In the equilibrium regime, the system can be efficiently described by an effective model in which the binders are traced out. Even in this case, the polymer display features that are different from those of a standard homopolymer interacting with two--body interactions. We then extend the effective model to deal with the case where binders cannot be regarded as in equilibrium and a new phenomenology appears, including local blobs in the polymer. Providing an effective description of the system can be useful in clarifying the fundamental mechanisms governing chromatin structuring.
\end{abstract}

\maketitle

\section{Introduction}

DNA--binding proteins, such as cohesin \cite{Ivanov2005}, HP1 \cite{James1986}, SATB1 \cite{Kohwi2012}, H-NS \cite{Dame2005} and many others, mediate the physical interactions between distal regions of chromatin. While some of them perform this task in a complex way, like the case of cohesin that consumes energy to extrude chromatin loops \cite{Goloborodko2016}, others work under conditions that are usually regarded as of near--equilibrium. For example, HP1 is regarded to mediate interactions in heterochromatin by binding to DNA and homodimerizing, with a mechanism that is weakly dependent on ATP consumption \cite{Eskeland2007}.

A simple model of the latter mechanism was developed by the Nicodemi group, who described the chromatin fiber as a polymer surrounded by diffusing beads that can bind to pairs of monomers of the polymer, mediating their mutual interaction \cite{Nicodemi2008}. This is referred to as string and binder switch (SBS) model. For homopolymers, they describe two phase transitions at increasing binder concentration or interaction energy \cite{Chiariello2016} from a coil to a disordered globule and then to a globule with an ordered arrangement of binders. Using a heteropolymeric model and multiple types of binders, the authors can reproduce chromatin folding into multiple domains \cite{Barbieri2012} observed experimentally \cite{Nora2012}.

On the other hand, several polymeric models were studied in the literature to describe profitably the conformational properties of the chromatin fiber, especially at large length scale, which are controlled by a direct interaction between their monomers \cite{Rosa2010,Mirny2011,Jost2014,Giorgetti2014,Michieletto16}. In all these models, the fact that the structuring of chromatin is controlled by binding proteins is treated implicitly.

A relevant question is whether one can trace out the degrees of freedom associated with the binders and build an effective model for the polymer. In fact, the motion of the chromatin fiber at the length scale of megabases, which is the most relevant for transcriptional control, takes place with a diffusion coefficient $D\approx 3\cdot 10^{-3}$ $\mu\text{m}^2/\text{s}$ \cite{Tiana2016} that is two orders of magnitude smaller than that of the proteins that mediate its interaction \cite{Schmiedeberg2004High-Cells}. One can then test the hypothesis that binders are at equilibrium during the motion of the fiber and calculate an effective potential for the polymeric chain that depends only on its degree of freedom. In the regime in which this approximation is valid, we expect that the system behaves as a homopolymer, displaying a transition between the standard coil and globular phases \cite{Grosberg1994}.

Using the analytical tools of statistical mechanics and molecular dynamics simulations of a simple polymeric model, we investigated both the regime in which the effective model applies and that in which binders cannot be regarded as at equilibrium and a new phenomenology appears.

The ability to simplify the description of the system while retaining its phenomenology is helpful in elucidating the basic mechanisms that control chromatin structuring in this case, and the behavior of physical systems in general.

\section{Binders display two equilibration regimes}

The starting point of this study are molecular dynamics simulations of the SBS model. We made use of a reference chain of $N_B=10^3$ identical monomers interacting with the potential
\begin{equation}
    U=U_{chain}+U_{HC}^B+U_{HC}^b+U_{LJ}^{Bb},
\end{equation}
where $U_{chain}=\frac{k}{2}\sum_i(|R_{i+1}-R_i|-a)^2$ is a harmonic potential that describes the polymeric bonds (setting $k=10^2$ and $a=1$, in arbitrary units), $U_{HC}^B=4\sum_I[\sigma^{12}/|R_I-R_J|^{12}-1/4]$ and $U_{HC}^b=4\sum_i[\sigma^{12}/|r_i-r_j|^{12}-1/4]$ are hard--core potentials on the beads of the chain and on the binders, respectively, and $U_{LJ}^{Bb}=4\epsilon\sum[\sigma^{12}/|R_I-r_j|^{12}-\sigma^6/|R_I-r_j|^6]$ is the attractive potential between the beads and the binders. Uppercase letters refer to the polymeric chain, lowercase letters to the binders.

The temperature is set to $T=1$ (in energy units, also setting Boltzmann's constant to 1) and the volume $V=60^3$ is defined by elastic boundary conditions. All masses are set to unity.
The friction coefficients are $\gamma_B=100$ ad $\gamma_b=1$, to reproduce the experimental ratio between diffusion coefficients (see above). We simulated the Langevin dynamics of the system with a timestep $\Delta t=10^{-3}$ for $10^9$ steps. Comparing the calculated and the experimental diffusion coefficient for a bead of the chain and assuming that $a=10^2$ nm, then a time unit of the model is $\approx 10^{-1}$ s, so each simulation represents a time span of the order of $10^5$ s $\sim 10^2$ h, which is larger than the duration of the cell cycle of mammals.
The simulations are performed with Lammps \cite{Thompson22},  varying the number of beads $n_b$ and $\epsilon$ that define the SBS model. 

\begin{figure}
    \centering
    \includegraphics[width=\linewidth]{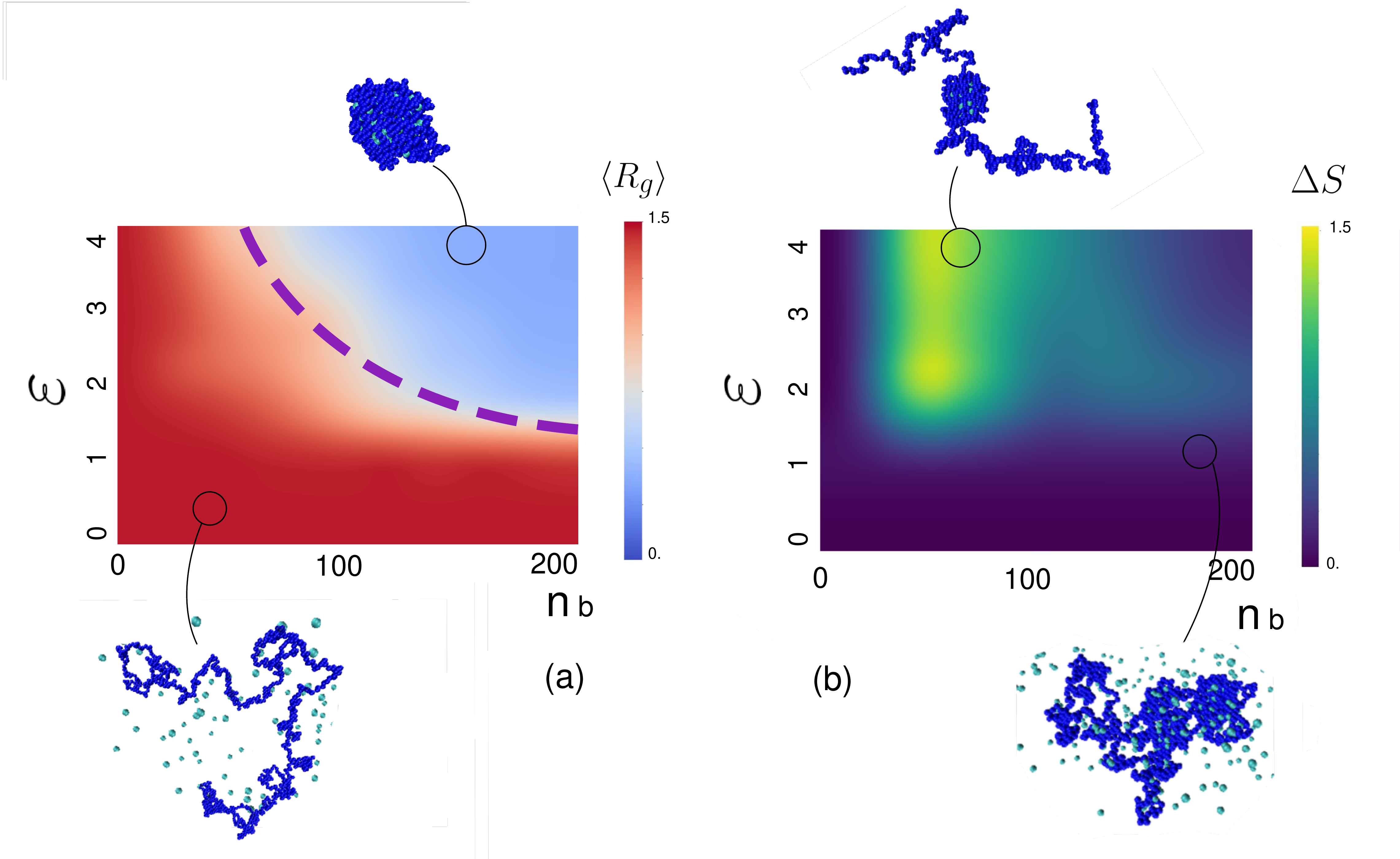}
    \caption{The values of the average gyration radius $\langle R_g\rangle$ normalized to $R_0=aN^{1/2}$ and of the order parameter $\Delta S$ that quantifies the localization of globules along the chain. The dashed line indicate the region in which the radius of gyration assumes the value of an ideal chain. Some representative snapshots of the system are displayed close to the plots.}
    \label{fig:sbs}
\end{figure}

The average gyration radius $\langle R_g\rangle$ seemingly displays a transition between globular and coil states of the chain along a curved line in the parameter space defined by the number of beads $n_b$ and the depth $\epsilon$ of the energy well (Fig. \ref{fig:sbs}a). The region corresponding to the ideal--chain behavior $\langle R_g\rangle=aN^{1/2}$ can be fitted with the curve 
\begin{equation}
\epsilon=\epsilon'-k\cdot\log\left[\frac{n_b}{n_b+k_b}\right]    
\end{equation}
with $\epsilon'=0.884$, $k=4.009$ and $k_b=36.415$ (dashed curve in Fig. \ref{fig:sbs}). The rationale of trying a fit with such a logarithm function is that it describes the loss of entropy upon binding in a two--state system.

However, while in the low--$\epsilon$ part of the parameter space the conformations of the chain look like the standard coil state of homopolymers (Fig. \ref{fig:sbs}b), in the rest of the parameter space it displays blobs that are not expected in the case of a homopolymer (Fig. \ref{fig:sbs}c; see also ref. \cite{Barbieri2012}).

The formation of localized blobs, involving only specific segments of the polymer, is a spontaneous breaking of the translational symmetry along the chain. It can be quantified by the Shannon entropy 
\begin{equation}
    \Delta S = S_{max}+\sum_I f_I\log f_I
\end{equation}
associated with the distribution of contacts $f_I=\langle\sum_j \theta(|R_I-r_j|<d_0) / \sum_{Ij} \theta(|R_I-r_j|<d_0) \rangle$ with binders along the chain; here, $\theta$ is a step function and $S_{max}=\log N_B$. The order parameter $\Delta S$ ranges from $0$ when the contacts are uniformly distributed along the chain to $S_{max}$ when they are localized in blobs. A non--uniform distribution of contacts ($\Delta S>0$) suggesting the presence of blobs appear stably only at large values of $\epsilon$. Although multiple blobs can appear in the early stages of simulations, they always coalesce into a single blob.

Since $\gamma_b\ll\gamma_B$, one can expect that the binders move much faster than the polymer and are in equilibrium as the polymer moves. This hypothesis could be valid at small $\epsilon$, where the binders can bind, unbind and diffuse often enough to make contacts with a uniform probability along the chain, but it is not compatible with the presence of stable blobs.
Thus, we have studied the degree of equilibration of the beads by repeating five independent simulations for each set of the parameters $\epsilon$ and $n_b$, calculating the average energy of the binders $U_{LJ}^{Bb}+U_{HC}^b$ in each simulation, and finally calculating the root mean square difference $\sigma$ among the average energies. If the binders are equilibrated within the simulation time, we expect that $\sigma$ is small, at least comparable to the thermal fluctuations within each simulation (that are at least of the order of $10^{-1}$ energy units). The quantity $\sigma$ can be compared with the standard deviation $\sigma_T$ due to thermal motion, calculated from the average of the variance over the five simulations. 

\begin{figure}
    \centering
    \includegraphics[width=\linewidth]{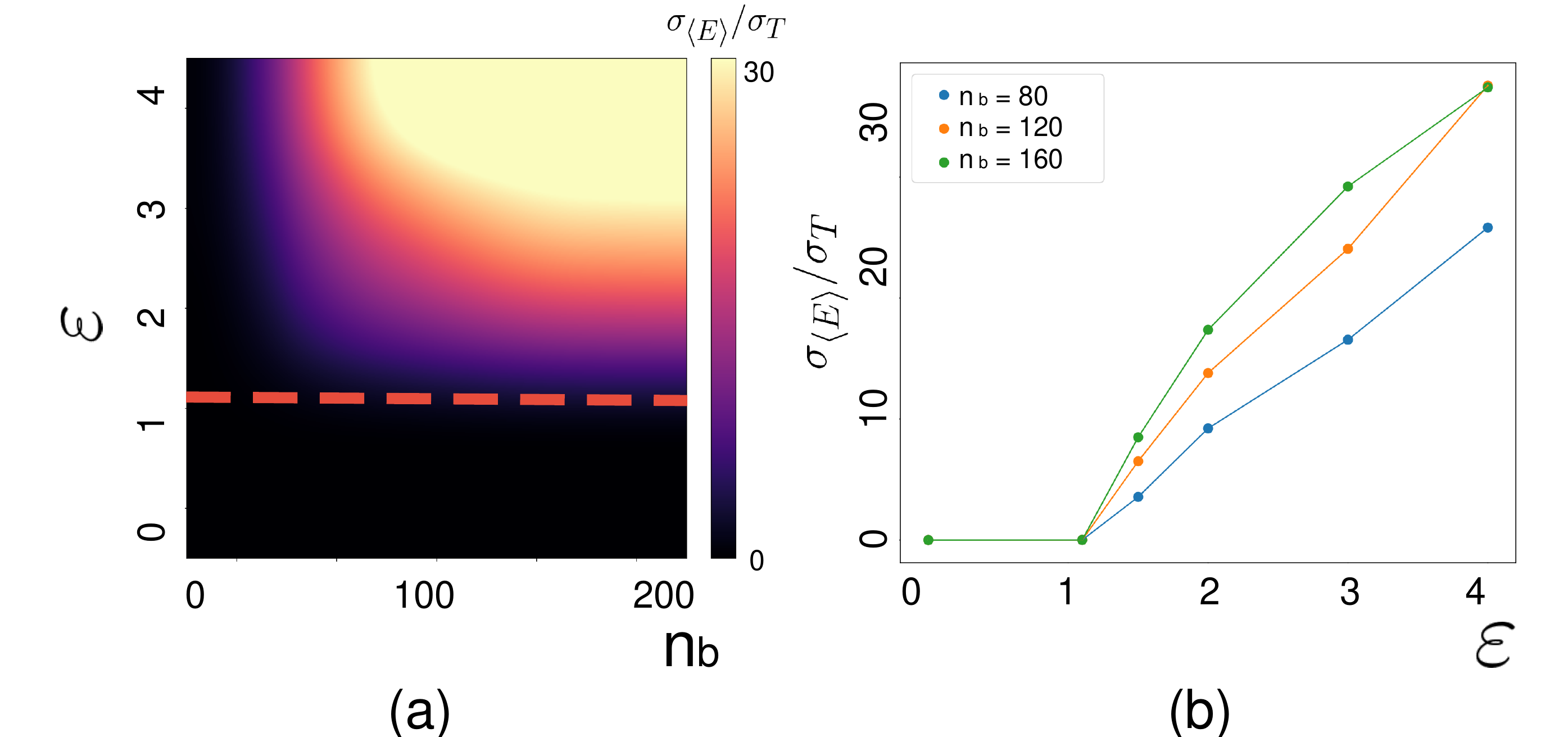}
    \caption{The root mean square difference $\sigma$ among the average energies $U_{LJ}^{Bb}+U_{HC}^b$ of the beads in five independent simulations, normalized to the standard deviation $\sigma_T$ due to thermal fluctuations. The dashed line indicates approximately where $\sigma$ starts to be larger than zero.}
    \label{fig:fluctdiss}
\end{figure}

The value of $\sigma/\sigma_T$ (Fig. \ref{fig:fluctdiss}) is approximately zero for $\epsilon\lesssim 1.2$ independently of $n_b$. Consequently, for larger values of $\epsilon$, binders cannot be at equilibrium. Although the requirement $\sigma\approx 0$ is only a necessary condition for equilibration of the binders, the fact that $\gamma_b\ll\gamma_B$ suggests that in the region $\epsilon<1.2$ they can move so faster than the chain that can be regarded as in equilibrium. The small value of $\Delta S$ in this region (Fig. \ref{fig:sbs}) supports this suggestion.

\section{The effective potential for the equilibrated regime}
\label{sect:effective}

We first developed an effective potential for the regime in which binders can be regarded as at equilibrium. For this purpose, we approximated the interaction energy between the binders and the monomers of the chain as a spherical--well potential
\begin{equation}
    U(\vR,\vvr)=-\epsilon\sum_{J=1}^{N_B}\sum_{i=1}^{n_b}\theta(|\vR_J-\vvr_i|<d_0),
    \label{eq:u0}
\end{equation}
where $\vR$ are the coordinates of the beads of the polymer, indexed by $i$, $\vvr$ are the coordinates of the floating binders, indexed by $J$, $\theta$ is a step function that is 1 if its argument is true and zero otherwise, $d_0$ is the interaction range and $\epsilon$ defines its strength. The binders are assumed not to interact with each other, and the same is true for the monomers of the chain.

The partial partition function, assuming that the $\vvr$ are at equilibrium for a given choice of $\vR$, is
\begin{equation}
    Z(\vR)=\int d^3r_1\dots d^3r_n \prod_{i=1}^{n_b}\exp\left[\beta\epsilon\sum_{J=1}^{N_B}\theta(|\vR_J-\vvr_i|<d_0)\right],
\end{equation}
where $\beta$ is the inverse temperature and we set Boltzmann's constant to unity. The free energy of the polymer is $F(\vR)=-T\log Z(\vR)$. Summing and subtracting $1$ to the integrand and realizing that the integrals factorizes over the binders, the free energy can be written as
\begin{equation}
    F(\vR)=-Tn_b\log\left[ V+\int d^3r \left( e^{\beta\epsilon\sum_J\theta(|\vR_J-\vvr_i|<d_0)}-1 \right)  \right].
\end{equation}
Since the step function that defines the integrand is piece--wise constant, we can integrate the equation above obtaining
\begin{equation}
    F(\vR)=-Tn_b\log\left[ V+ \sum_k V_k(\vR) \left(e^{\beta\epsilon k}-1\right) \right],
    \label{eq:f}
\end{equation}
where $V_k(\vR)$ is the volume of the intersection of the interaction volume of $k$--plets of beads of the polymer, that is
\begin{align}
    V_k(\vR)&\equiv \frac{1}{k!}\sum_{R_1,R_2,...,R_k}\int d^3r\, \theta(|\vR_1-\vvr|<d_0)\cdot \nonumber\\
    &\cdot\theta(|\vR_2-\vvr|<d_0)\dots\theta(|\vR_k-\vvr|<d_0).
\end{align}

The free energy given by Eq. (\ref{eq:f}) can be regarded as an effective energy for the polymeric chain. It depends linearly on the number of binders, it is non--additive because of the logarithm and has a many--body character because of the sum on $k$. The $k$th--order term is attractive if $k>T/\epsilon$, otherwise it is repulsive. This means that all terms are attractive if $T<\epsilon$. 

The dependence of the effective potential on the mutual distance between the beads is in general nontrivial. For the $2$--body part it is
\begin{align}
    V_2(\vR)=&\frac{\pi}{12}\sum_{I<J}(4d_0+|\vR_I-\vR_J|)(2d_0-|\vR_I-\vR_J|)^2 \cdot \nonumber\\
    &\cdot\theta(|\vR_I-\vR_J|<d_0),
\end{align}
and other expressions that depend explicitly on the mutual distances among beads exist up to $k=6$ \cite{Chkhartishvili2015}. Anyway, the resulting effective forces
\begin{equation}
    \mathbf{f}=-\mathbf{\nabla} F(\vR)
    \label{eq:force}
\end{equation}
are not two--body even neglecting in Eq. (\ref{eq:f}) the terms $k>2$.

The effective energy defined by Eq. (\ref{eq:f}) has an important drawback, namely that the absolute value of the energetic part encoded by the exponential grows with $k$, making the higher--order terms potentially the most important. However, the derivation done so far starting from Eq. (\ref{eq:u0}) does not include a hard--core repulsion that would make the description of the chain more realistic. The effects of a repulsive term in the initial potential are to decrease the volume of intersection among the interaction spheres and to prevent the overlap of multiple spheres. In the limit in which the hard--core radius becomes equal to $d_0$, only the term $k=2$ is relevant. In other words, one can truncate the sum over $k$ not because the energetic terms are a decreasing series at increasing $k$, but because the $V_k$ are exactly zero.

Due to the presence of the logarithm in Eq. (\ref{eq:f}), the effective energy cannot be written in general as the sum over pairs of monomers, even considering only the case $k=2$; thus it does not qualify as a two--body interaction. Only in the limit of large temperature (or large volume $V$) it approximates to
\begin{equation}
    F(\vR)=-T\frac{n_b}{V}\sum_k V_k(\vR) \left(e^{\beta\epsilon k}-1\right),
\end{equation}
and then becomes a series over $k$--body interactions. Under this approximation, the effective energy is a function of the concentration of binders and not on their copy number.

\section{The effective model recapitulates the properties of the polymer at equilibrium}

We compared the simulations of the dynamics of a polymer described with the SBS model with those obtained with the effective model in the two--body approximation, that is neglecting the terms with $k>2$ in Eq. (\ref{eq:f}). We chose the radius of interaction of the effective model $d_0=1.15$, corresponding to the distance at which the Lennard---Jones energy of the SBS model display an energy that is 2/3 of the minimum ($\epsilon=1.1$). 

\begin{figure}
    \includegraphics[width=\linewidth]{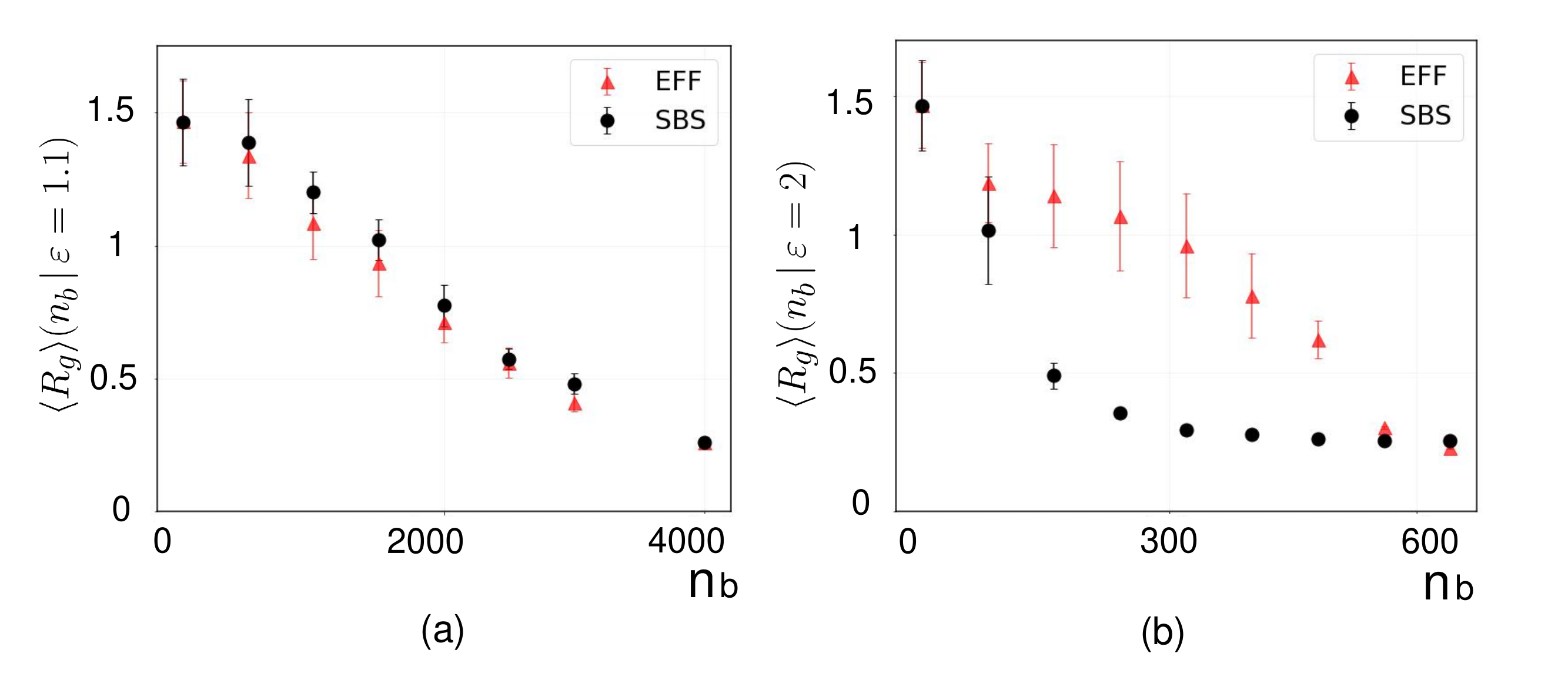}
    \caption{Comparison of the radius of gyration $R_g$ obtained from the SBS and from the effective model at two different values of $\epsilon$. The error bars indicate the standard deviation.}
    \label{fig:eff}
\end{figure}

The average radius of gyration of the chain simulated with the effective model (left panel in Fig. \ref{fig:eff}) is barely distinguishable (p--value=0.64) from that of the SBS model in the regime in which the beads are equilibrated, e.g. at $\epsilon=1.1$ (cf. Fig. \ref{fig:fluctdiss}). This result suggests that not only the effective model describes correctly the geometrical properties of the SBS system, but also the approximation we did neglecting the terms with $k>2$ in Eq. (\ref{eq:f}) is reasonable.

On the other hand, at larger values of $\epsilon$, the radius of gyration obtained from the effective model is very different from that of the SBS model (p--value=0.04; right panel in Fig. \ref{fig:eff}). Not unexpectedly, in this regime the effective model fails because the binders are not equilibrated (Fig. \ref{fig:fluctdiss}) and so the main hypothesis underlying the effective model is not valid.

\begin{figure}
    \includegraphics[width=\linewidth]{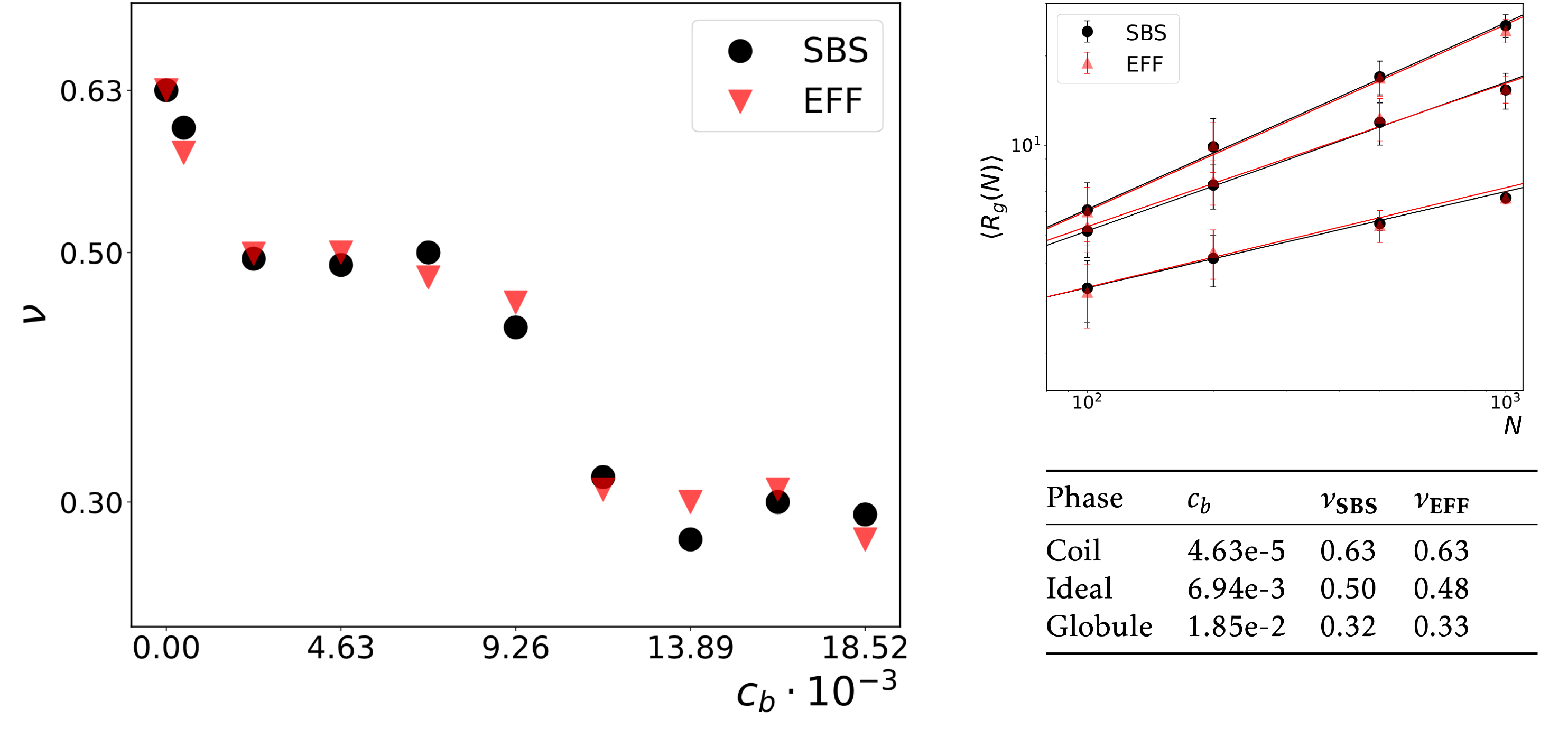}
    \caption{The scaling of the average radius of gyration with the length of the polymer (right panel) for three different concentration $c_b$ of binders (cf. table below the panel) at $\epsilon=1.1$. The scaling exponent $\nu$ as a function of $c_b$.}
    \label{fig:n}
\end{figure}

In the equilibrated regime, the polymer displays the scaling properties of the standard coil and globule phases with both models (Fig. \ref{fig:n} shows the case $\epsilon=1.1$). The scaling of the radius of gyration with the length $N$ of the polymer was studied at fixed values of the concentration $c_b=n_b/V$ of binders, varying the volume to accommodate polymers of different length ($V=216\cdot N$). The scaling exponent $\nu$ of the SBS and of the effective models are indistinguishable from each other (right panel in Fig. \ref{fig:n}).  

The exponent $\nu$ starts at $c_b=0$ from the value $\approx 3/5$ typical of a random coil (left panel in Fig. \ref{fig:n}). Alreasy at low concentration of binders, $\nu$ drops to the ideal--chain value $\approx 1/2$ and makes a wide plateau that is non apparent in standard homopolymeric models with two--body interactions. Finally, $\nu$ drops again to $\approx 1/3$, corresponding to the globlar phase.


\section{Conformational properties of the SBS model in non--equilibrium regime} 
\label{sect:oute}

\begin{figure}
    \includegraphics[width=\linewidth]{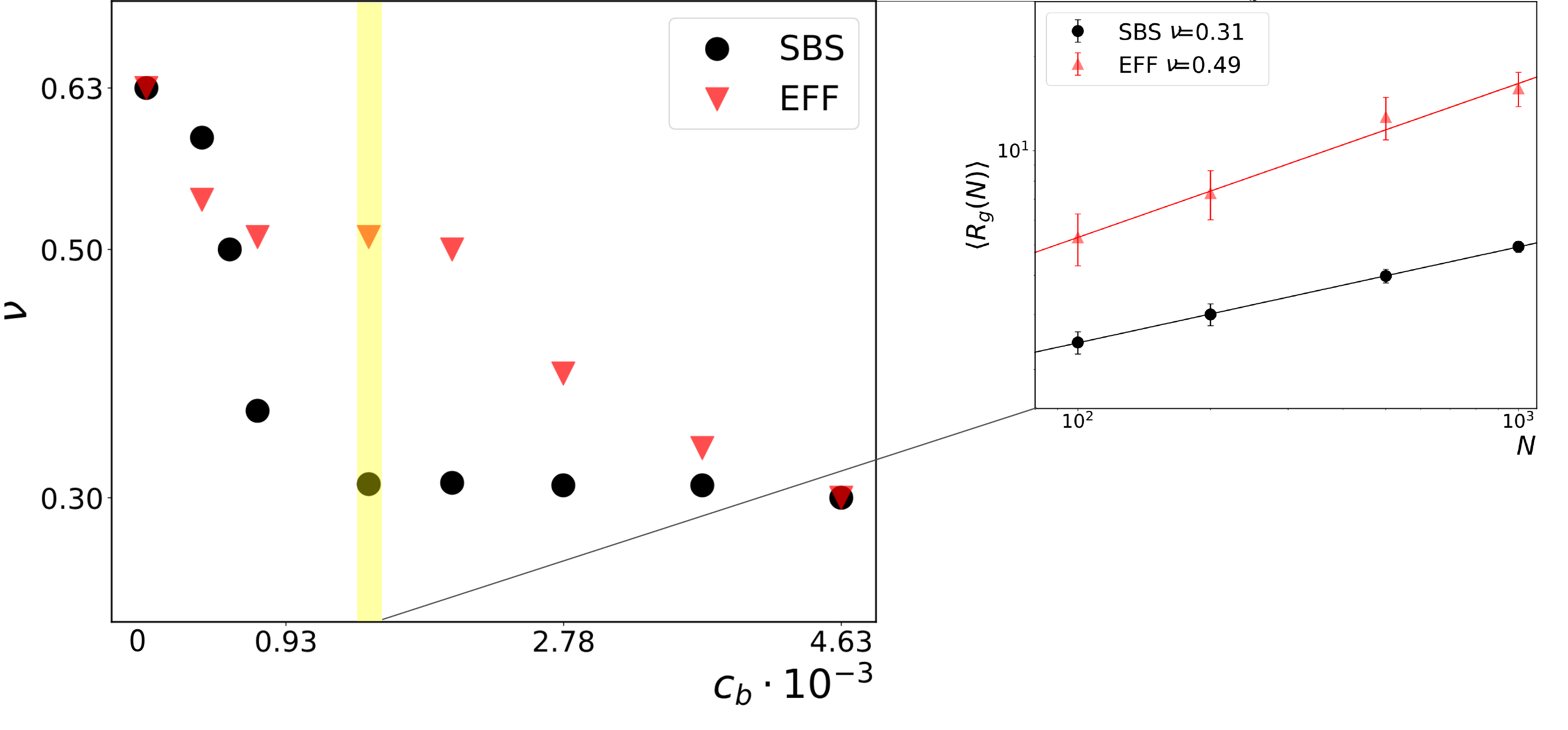}
    \caption{The scaling of the radius of gyration with the length of the polymer (right panel) and the scaling exponent $\nu$ for $\epsilon=2.0$, corresponding to the non--equilibrium regime.}
    \label{fig:nne}
\end{figure}

Apparently, the conformational properties of the polymer are more similar to standard homopolymers in the non--equilibrium regime. In fact, the scaling exponent of the SBS model (Fig. \ref{fig:nne}) drops from the value typical of random coils $\nu \approx 3/5$ to that typical of globules ($\nu\approx 1/3$), without displaying any plateau at the ideal--chain value ($\nu=1/2$). However, the effective model derived under the equilibrium hypothesis gives values of $\nu$ that are significantly different from those of the SBS model.

We already know that a distinctive feature of the non--equilibrium regime is the uneven distribution of contacts in the system (cf. Fig. 1, right panel), that corresponds to the presence of blobs. We define the relative size $f_B$ of the blob as the fraction of beads of the polymer belonging to the largest cluster of mutually--interacting beads, that is
\begin{equation}
   f_B =\frac{1}{N_B} \max_{\gamma} |\gamma|
\end{equation}
where $\gamma$ labels the clusters, defined as subsets of beads connected in a graph by mutual distances below $d_0$; $|\gamma|$ denotes the number of nodes of a cluster.

\begin{figure}
    \centering
    \includegraphics[width=\linewidth]{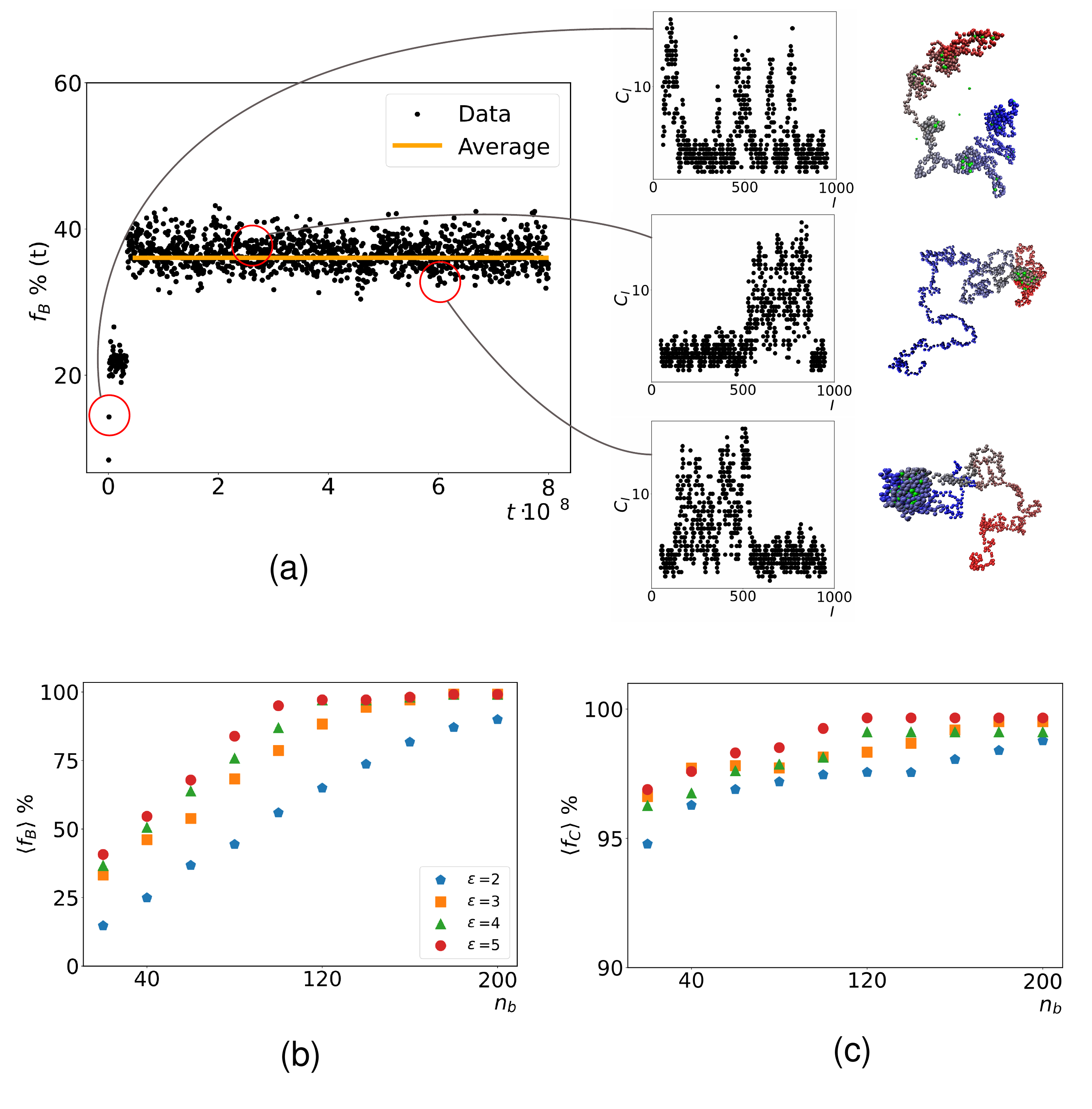}
    \caption{(a) The time dependence of the fraction $f_b$ of beads participating to a blob in a sample simulation starting from a random coil with $n_b=60$ and $\epsilon=2$. Three snapshots taken from the simulation and the associated number of contacts per bead are displayed on the side. The average of (b) $f_B$ and (c) of $f_C$  as a function of the number of binder $n_b$ in the SBS model, for different values of $\epsilon$. }
    \label{fig:blobs}
\end{figure}

Simulations of the SBS model starting from a random coil reach a stationary distribution of $f_B$ and fluctuate weakly around it (Fig. \ref{fig:blobs}a). The average stationary fraction $\langle f_B\rangle$ of beads of the polymer in the blob  increases as a function of the concentration of binders (Fig. \ref{fig:blobs}b) from 0, reaching 1 when the chain collapses to a globule (cf. Fig. \ref{fig:nne}). The value of $\langle f_B\rangle$ increases with $\epsilon$, saturating at $\epsilon\approx 5$. At these large values of $\epsilon\gg kT$ all the binders are bound to the polymer and can only exchange the interacting bead of the chain, moving along it; consequently, increasing $\epsilon$ has no effect. 

As the beads participating to the blob becomes stationary in number, they also tend to group along the chain (Fig. \ref{fig:blobs}a, left insets). We quantified this effect counting the fraction of  beads that are consecutive in the largest cluster $\gamma^*$
\begin{equation}
    f_C = \frac{1}{|\gamma^*|}\sum_{I=1}^{N_B-1}\theta(I\in\gamma^*)\,\theta(I+1\in\gamma^*)
\end{equation}
The average $\langle f_C \rangle$ is always larger than $0.95$ (Fig. \ref{fig:blobs}c), indicating that the blobs are typically made of consecutive beads.
This fact suggests that the entropy of the polymer is largest when they constrain as few beads as possible, so they rapidly come to form a blob in which few beads share as many binders as possible, compatibly with the mutual hard--core repulsion.

Moreover, although the number of beads of the polymer participating to the blob is quite constant, its position can shift along the chain (Fig. \ref{fig:blobs}a, left inset). This is a consequence of the fact that the upraise of the blob is a spontaneous breaking of the translational symmetry of the potential function of the system that fluctuations tend to compensate.

\section{Extension of the effective model to the non--equilibrium regime}

The effective model introduced in Sect. \ref{sect:effective} fails at large values of $\epsilon$ because it is no longer true that the binders can be regarded as in equilibrium. We then looked for the simplest modification of the effective forces of Eq. (\ref{eq:force}) to reproduce the phenomenology described in Sect. \ref{sect:oute}. 

For this purpose, we maintained the definition of the effective energy of Eq. (\ref{eq:f}), but we rescaled the force acting on each bead of the chain by an effective local number of free binders
\begin{equation}
    n_b(I)\equiv \left[ \alpha\, n_b-\beta\sum_{K,J\not\in\gamma(I)}\theta(|\vR_K-\vR_J|<d_0) \right]_+,
\end{equation}
where $\gamma(I)$ labels the clusters of mutually interacting beads containing bead $I$ (i.e., a subset of beads connected by distances below $d_0$ to bead $I$), $\alpha$ and $\beta$ are two parameters of the effective model, describing the effective number of binders per nominal binder and the average number of binders per contact, respectively; the square brackets indicate a rectifying function. Consequently, the force acting on the $I$th bead is
\begin{equation}
    \mathbf{f_I}=-\frac{n_b(I)}{n_b}\mathbf{\nabla_I} F(\vR).
    \label{eq:force}
\end{equation}
The goal of this renormalization is to account for the fact that if many contacts are formed elsewhere, and thus binders are effectively sequestered, less of them are available to strengthen the interactions of a given bead.
Note that we have modified the forces, and not the effective potential, with a conformational--dependent parameter. This is a way to introduce non--Markovianity in the system, because $n_b(I)$  accounts for the slowly--varying state of the system and is not differentiated as if it reweighted the effective energy. The fact that the reweighting is the same for all beads of the same cluster enforces the third Newton's law.
 
\begin{figure}
    \centering
    \includegraphics[width=\linewidth]{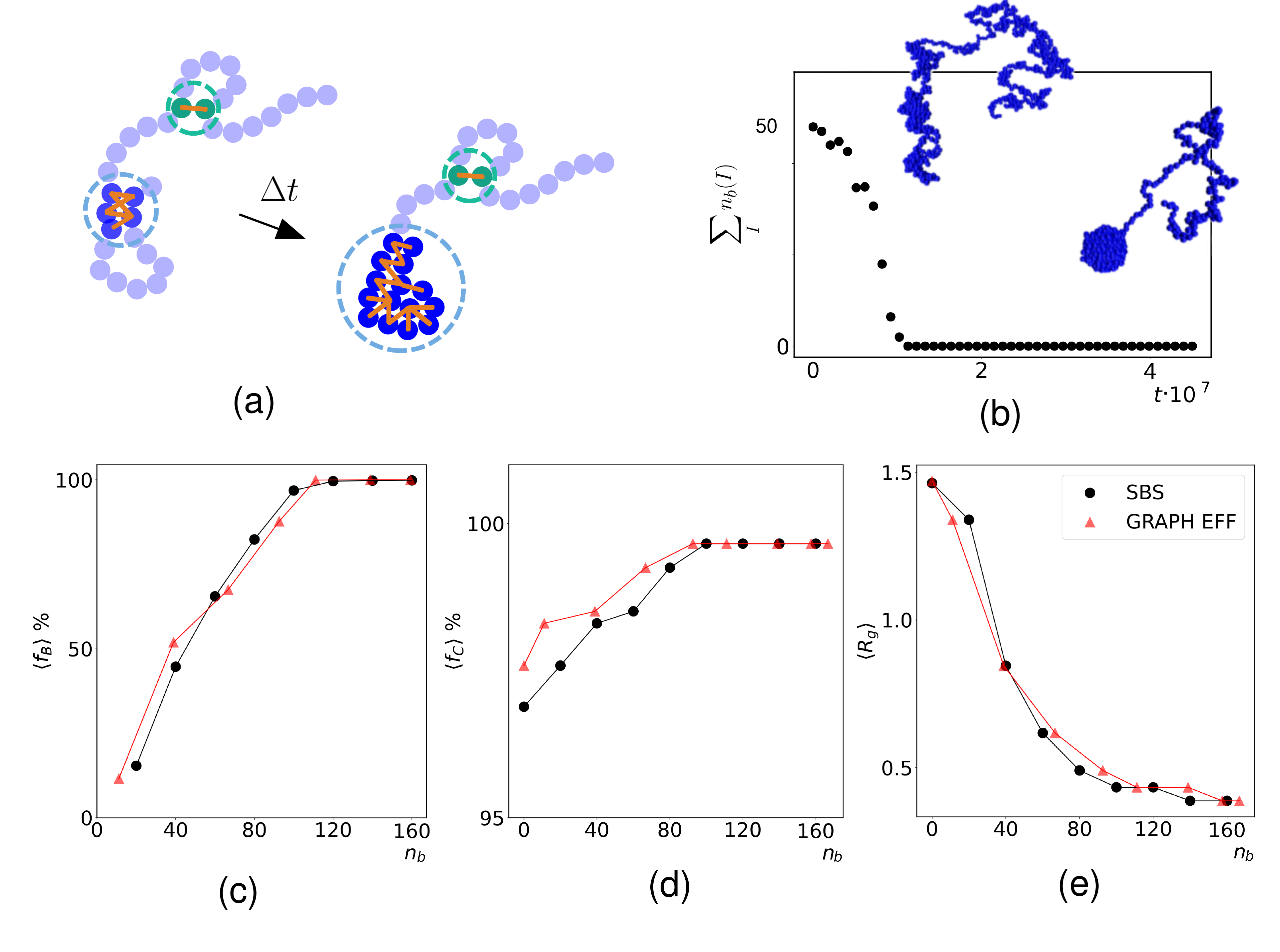}
    \caption{(a) A sketch of the out--of--equilibrium effective model that penalizes the strength of the interactions in a cluster when other clusters are present. (b) The effective number of free binders $\sum_I n_b(I)$ along a sample trajectory for the effective model with $\epsilon = 5$, $\epsilon=5$, $\alpha=0.02$ and $\beta=0.5$. (c) The average fraction of beads in the blob, (d) the average fraction of consecutive beads in the blob and (e) the radius of gyration for the effective model (in red) and for the SBS model (in black)}
    \label{fig:oute}
\end{figure}

Simulations of this model with $\epsilon=5$, $n_b=2500$, $\alpha=0.02$ and $\beta=0.5$ show that the effective number of free binders $\sum_I n_b(I)$ goes rapidly to zero while blobs get formed (Fig. \ref{fig:oute}b). The average size $\langle f_b\rangle$ of the blob and its degree of locality $\langle f_C\rangle$ along the chain of this effective model are similar to those of the SBS model (Fig \ref{fig:oute}c,d) for this choice of the parameters $\alpha$ and $\beta$. Also the radius of gyration is similar in the two models (Fig. \ref{fig:oute}e).

Thus, this model is able to reproduce the behavior of the SBS model in the non--equilibrium regime for a choice of the parameters $\alpha$ and $\beta$ that define it, taking 
implicitly into account the non--equilibrium dynamics of the binders and their finite number.

We also tested other models, like a Markovian one in which Eq. (\ref{eq:f}) is modified substituting the effctive number of free binders $n_b-\vartheta\sum_{IJ} \theta(|R_I-R_J|<d_0)$ to $n_b$, or another in which the effective number of free binders is evaluated at the time at which a contact is formed, and then kept fixed, thus giving rise to a non--Markovian dynamics. With none of these models we observed the stabilization of blobs as described in Sect. \ref{sect:oute}.

\section{Discussion and conclusions}

\begin{figure}
    \centering
    \includegraphics[width=\linewidth]{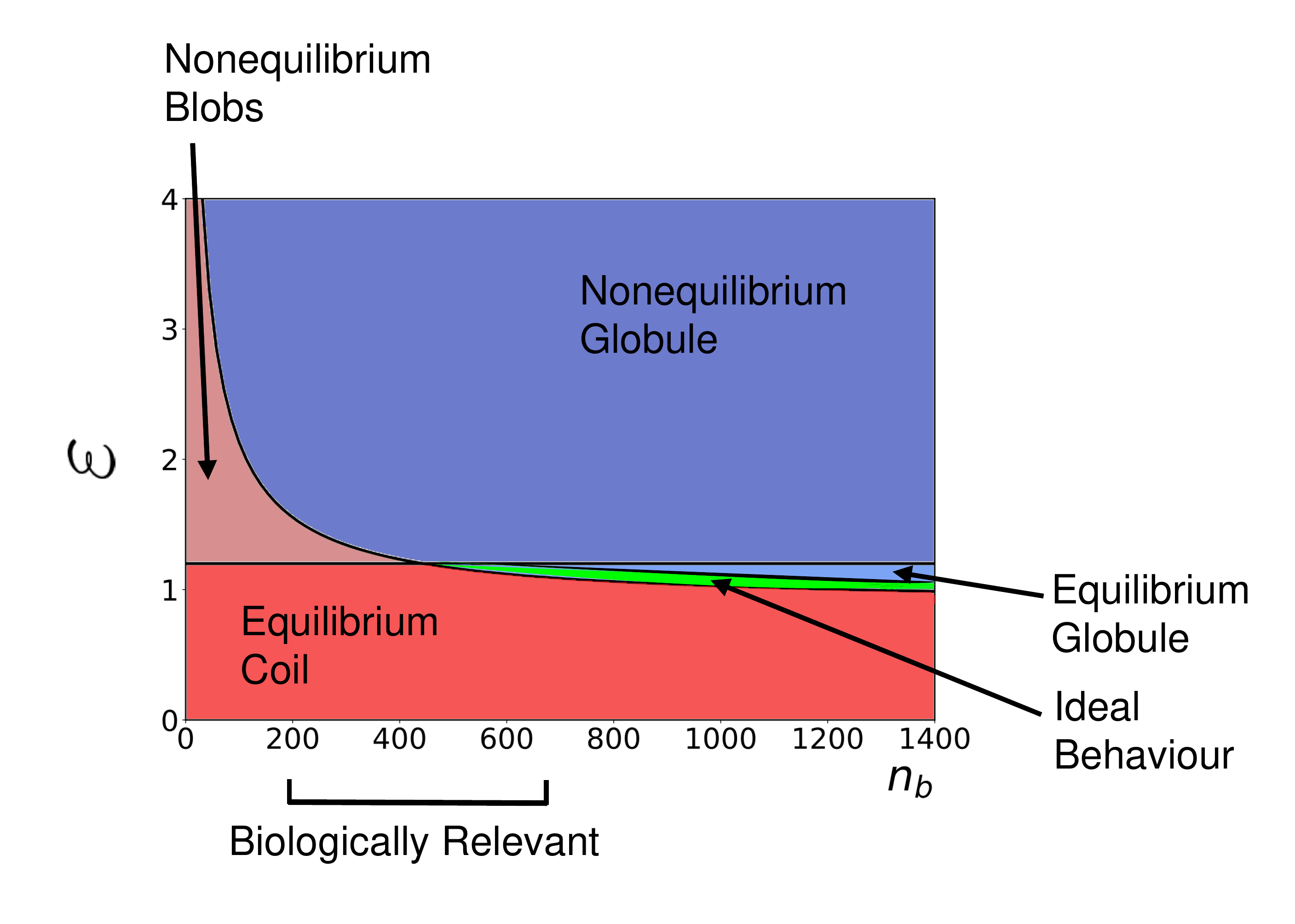}
    \caption{A sketch of the phase diagram of the system}
    \label{fig:pd}
\end{figure}

The simulation of a polymeric model whose interaction is mediated by fast--diffusing binders shows two different regimes as a function of the interaction strength of the binders. When this quantity is small, the binders can be regarded as in equilibrium and the polymer displays  three phases similar to standard homopolymers, namely coil, ideal and globule. However, the phase diagram (Fig. \ref{fig:pd}) seems more complex than the standard one. For example, there is a wide range of binder concentration in which the polymer display the same scaling law of the ideal polymer, but with varying gyration radii.

When the interaction strength is large, the binders cannot be regarded as in equilibrium as the polymer moves. Now the polymer can populate phases that display the sizes typical of coil, ideal chains and globules, but display localized blobs that cannot be stable in a standard homopolymer. Being the beads of the polymer model indistinguishable from each other, the blobs reflect a spontaneous breaking of the translational symmetry of the interaction potential along the chain. The observed motion of the blob along the chain thus plays the role of a Goldstone modes associated with such a symmetry breaking. In a more realistic description of chromatin, different loci can display varying propensities for binding proteins, the interaction can be not translationally symmetric and the Goldstone modes can disappear.

Structure factors of chromatin like Hp1 have residence times that range from 0.2 s in weak binding sites of euchromatin to 2 min in heterochromatin \cite{Muller2009}. This range coincides with the typical times associated with the motion of chromosomes, that ranges from $t = \ell^2/D\sim 1$~s on the scale of chromosomal domains ($\ell=10^2$ nm, $D=3\cdot 10^{-3}$ $\mu\text{m}^2/\text{s}$ \cite{Tiana2016}) to minutes for the whole chromosome ($\ell\sim 1$ $\mu\text{m}$). As a consequence, there is no separation of time scales between the motion of the binders and that of the polymer, and the binders can be either in the equilibrium or in the non--equilibrium regime according to the details of the specific loci of interest. Both regimes found in the present model study are then relevant for chromatin.

In the original description of the SBS homopolymeric model \cite{Chiariello2016}, the authors describe a phase diagram displaying a coil--globule transition and a transition between ordered and disordered binders. Now we can add some new elements to this description. First, the high--$\epsilon$ branch of the coil--globule transition and all the order--disorder transition occur in the out--of--equilibrium regime, and thus cannot be described by equilibrium statistical mechanics. Moreover, some non--trivial behavior that leads to the appearance of blobs occurs when the number of binders is small, smaller than that explored in the original work (starting around $n_b\approx 200$ for a polymer of $N=10^3$). This range is biologically relevant because the copy number of architectural proteins in chromosomal domains regarded at kbp--resolution is variable and can be much smaller \cite{Muller-Ott2014} than that regarded in that work.

The development of effective models that average out the degrees of freedom associated with binders can be helpful not only for  computational convenience, but also because they can highlight the physical mechanisms responsible for the observed phenomenology.

Even in the simpler effective model that regards binders at equilibrium, the effective interaction is not two--body. This is the main reason why this system display a phase diagram that is more complex than that predicted by the Flory theory of standard homopolymers.  

Increasing the interaction strength or decreasing the number of the binders, these cannot be regarded as in equilibrium because the time scale associated with their motion is not smaller than that associated with the motion of the polymer. According to the Mori--Zwanzig theory \cite{Zwanzig1961}, a dimensional reduction gives rise to a dynamics controlled by thermal noise and the effective energy is the negative gradient of the free energy only if the discarded degrees of freedom are fast with respect to those which are retained. This is not the case for the non--equilibrium regime, and in this case the Mori--Zwanzig theory predicts a non--Markovian dynamics of the reduced system. This is the reason why we could reproduce the phenomenology of the SBS model only modifying the forces in a non--Markovian way. The apperaence of stable blobs is then a consequence of such non--Markovianity.

\end{document}